 \documentstyle[twoside,fleqn,espcrc2,epsf]{article}

\newcommand{\AmS}{{\protect\the\textfont2
  A\kern-.1667em\lower.5ex\hbox{M}\kern-.125emS}}

\title{Dynamical ordering in the c-axis in 3D driven vortex
lattices}

\author{
	Alejandro B. Kolton, Daniel Dom\'{\i}nguez 
	\address{Centro At\'{o}mico Bariloche,\\ 8400 S. C. de Bariloche,
	Rio Negro, Argentina}		
        and 
        Niels Gr{\o}nbech-Jensen	
	\address{Department of Applied Science, University of California, \\
	Davis, California 95616, USA}
	\address{NERSC, Lawrence Berkeley National Laboratory,\\
        Berkeley, California 94720, USA}
	}
       
\begin{document}

\begin{abstract}
We present molecular dynamics simulations of driven vortices in layered
superconductors in the presence of an external homogeneous force and point
disorder. We use a model introduced by J.R.Clem for describing 3D vortex
lines as stacks of 2D pancake vortices where only magnetic interactions
are considered and the Josephson interlayer coupling is neglected. We
numerically evaluate the long-range magnetic interaction between pancake
vortices exactly. We analyze the vortex correlation along the field
direction on (c-axis). We find that above the critical current, in the
``plastic flow'' regime, pancakes are completely uncorrelated in the
c-direction. When increasing the current, there is an onset of correlation
along the c-axis at the transition from plastic flow
 to a moving smectic phase. This transition coincides with the peak in the
differential resistance.
\vspace{1pc}
\end{abstract}

\maketitle

\section{INTRODUCTION}
The prediction \cite{KV} of a {\it dynamical phase transition} upon increasing 
drive, from a fluidlike plastic flow regime \cite{plastico} to a 
coherently moving solid \cite{KV} in a moving vortex lattice has motivated many 
recent theoretical \cite{theo}, experimental 
\cite{bhatta,hellerq,pardo}, and 
simulation work \cite{2d,ryu,olson2,kolton,mingo,tdgl,3dxy}.    
In a previous work we have studied the dynamical regimes in the velocity-force
curve (voltage-current) in 2D thin films \cite{kolton} and found two
dynamical phase transitions above the critical force. The first transition, from a 
plastic flow regime to a smectic flow regime, is characterized by the 
simultaneous occurrence of a peak in differential resistance, isotropic low 
frequency voltage noise and maximum transverse diffusion. 
The second transition, from a smectic flow regime to a frozen transverse
solid, is a freezing transition in the transverse direction where transverse 
diffusion vanishes abruptly and the Hall noise drops many orders of magnitude.
In other 2D simulations the peak in differential
resistance was found to coincide with the onset of orientational order \cite{ryu} 
and a maximum number of defects \cite{olson2}. Experimentally, the position of 
this peak was taken by Hellerqvist {\it et al.} \cite{hellerq} as an 
indication of a dynamical phase transition. 
In this paper we show that in driven 3D layered superconductors, the peak in 
differential resistance also 
coincides with the onset of correlation along the c-axis.    
\section{MODEL}
Simulations of vortices in 3D layered superconductors have been done previously using 
time--dependent Ginzburg--Landau--Lawrence--Doniach equations (two layers), 
\cite{tdgl}, the 3D XY model \cite{3dxy} and Langevin dynamics of interacting 
particles \cite{reefman,srld}.  
Here we study the motion of pancakes vortices in a layered superconductor with
disorder, with an applied magnetic field in the c-direction and with an external 
homogeneous current in the layers (ab-planes). We use a model introduced by 
J.R. Clem for a layered superconductor with vortices in the limit of zero 
Josephson-coupling between 
layers \cite{clem}. 
We simulate a stack of equally spaced superconducting layers with interlayer periodicity $s$, 
each layer containing the same number of pancake vortices. The equation of motion for a 
pancake located in position ${\bf R_{i}}=({\bf r_{i}},z_i)=(x_{i},y_{i},n_i s)$ 
(with z-axis in c-direction) is:
\begin{equation}
\eta \frac{ d{\bf r_i}}{dt} = \sum_{j\not= i}{\bf F_v}(\rho_{ij},z_{ij})
+\sum_p{\bf F_p}(\rho_{ip}) + \bf{F} \; ,
\end{equation}
where $\rho_{ij}=|{\bf r}_i-{\bf r}_j|$ and $z_{ij}=|z_i-z_j|$ are 
the in-plane and inter-plane distance between pancakes $i,j$, $r_{ip}=
|{\bf r}_i-{\bf r}_p|$ is the in-plane distance between the vortex $i$ and a pinning 
site at ${\bf R_p}=({\bf r}_p,z_i)$ (pancakes interact only with pinning centers within the same 
layer),
 $\eta$ is the Bardeen-Stephen friction,
 and ${\bf F}=\frac{\Phi_0}{c}{\bf J}\times{\bf z}$
is the driving force due to an applied in-plane current density ${\bf J}$. 
We consider a random uniform distribution of attractive pinning centers in each layer with 
${\bf F_p}=-A_p e^{-(r/r_p)^2} {\bf r}/r_p^2$, where $r_p$
is the pinning range.
The magnetic interaction between pancakes ${\bf F}_v(\rho,z)=F_{\rho}(\rho,z) \hat{r}$ is
given by:
\begin{equation}
F_{\rho}(\rho,0)=(\phi_0^2 / 4 \pi^2 \Lambda \rho)[1-(\lambda_{\parallel}
/\Lambda)(1-e^{-\rho / \lambda_{\parallel}})]
\end{equation}
\begin{equation}
F_{\rho}(\rho,z)=(\phi_0^2 \lambda_{\parallel}/ 4 \pi^2 \Lambda^2 \rho)
[e^{-z/\lambda_{\parallel}}-e^{-R/\lambda_{\parallel}}] \; .
\end{equation}
Here, $R=\sqrt{z^2+\rho^2}$, $\lambda_{\parallel}$ is the penetration length parallel to
the layers, and $\Lambda=2 \lambda_{\parallel}^2/s$ is the 2D thin-film screening
length. A analogous model to Eqs.\ (2-3) was used in \cite{reefman}. 
We normalize length scales by $\lambda_{\parallel}$, energy scales by 
$A_v=\phi_0^2 / 4 \pi^2 \Lambda$, 
and time is normalized by 
$\tau=\eta \lambda_{\parallel}^2/A_v$. We consider $N_v$ pancake vortices and $N_p$ pinning
centers per layer in $N_l$ rectangular layers of size $L_x\times L_y$, 
and the normalized vortex density is $n_v=N_v\ \lambda_{\parallel}^2 /
 L_xL_y=B\ \lambda_{\parallel}^2 /\Phi_0$.
Moving pancake vortices induce a total electric field  ${\bf
E}=\frac{B}{c}{\bf v}\times{\bf z}$, with ${\bf v}=\frac{1}{N_v N_l}\sum_i 
{\bf v}_i$.
We study the dynamical regimes in the velocity-force 
curve at $T=0$, solving Eq.\ (1) for increasing values of ${\bf F}=F{\bf y}$. 
We consider a constant vortex density $n_v=0.1$ in
$N_l=5$ layers with $L_x/L_y=\sqrt{3}/2$, $s=0.01$, and 
$N_v=36$ pancake vortices per layer. 
We take a pinning range of $r_p=0.2$, pinning strengh of $A_p/A_v=0.2$,
with a density of pinning centers $n_p=0.65$ on each layer. We use periodic boundary
conditions in all directions and the periodic long-range in-plane interaction is dealt
with exactly using an exact and fast converging sum \cite{log}. 
The equations are integrated with a time step of $\Delta t=0.01\tau$ and averages are
evaluated in $16384$ integration steps after $2000$ iterations for 
equilibration (when the total energy reaches a stationary value). 
Each simulation is started at $F=0$ with an ordered
triangular vortex lattice (perfectly correlated in c-direction) and slowly increasing 
the force in steps of $\Delta F= 0.1$ up to values as high as $F=8$.

\section{RESULTS}

We start by looking at the vortex trajectories in the steady state 
phases.
In Figure 1(a-b) we show a top view snapshot of the instantaneous pancake 
configuration for two typical values of $F$. In Figure 2(a-b) we show the vortex 
trajectories $\{ {\bf R}_i(t)\}$ for the same two typical values of $F$ by 
plotting all the positions of the pancakes in all the layers for all the time 
iteration steps. In Fig.\ 3(a) we plot  the average vortex velocity 
$V=\langle V_y(t)\rangle=\langle\frac{1}{N_v}\sum_i \frac{dy_i}{dt}\rangle$,
in the direction of the force as a function of $F$ and its corresponding
derivative $dV/dF$. We also study the pair distribution function:
\begin{equation} 
g(\rho,n)=\frac{1}{N_p N_l}\langle \sum_{i<j}
\delta(\rho-\rho_{ij}) \delta(ns-z_{ij})  \rangle.
\end{equation}
In Fig.\ 3(b) we plot a correlation parameter along c-axis defined as:
\begin{equation} 
C^n_z=\lim_{\rho \rightarrow 0} g(\rho,n) \; ,
\end{equation}
as a function of $F$ for $n=1,2$.  
Below a critical force, $F_c \approx 0.4$, all the pancakes are
pinned and there is no motion. At the characteristic force,
$F_p \approx 0.8$, we observe a peak in the differential resistance. 
At $F_c$ pancake vortices start to move in a few channels 
, as was also seen in 2D vortex simulations \cite{plastico}. A typical 
situation is  shown in Fig.\ 2(a). In this plastic flow regime 
we observe that the motion is completely uncorrelated along c-direction, 
with $C^n_z \approx 0$ for $F_c< F <F_p$ 
as shown in Fig.\ 3(b). In Fig.\ 1(a) and 2(a) we see that this situation corresponds to 
a disordered configuration of pancakes and to an uncorrelated structure of plastic 
channels along c-axis. At $F_p$ there is an onset of correlation along the 
c-axis and pancakes vortices start to align forming well defined stacks or 
vortex lines. This onset of c-axis correlation corresponds to the transition 
from plastic flow to a moving smectic phase
(a complete discussion of the translational 
as well as temporal order will be discussed elsewhere \cite{veremos}). For $F > F_p$ we 
observe that the structure of smectic channels is very correlated in the c-direction. 
\begin{figure}[htb]
\centerline{\epsfxsize=8cm \epsfbox{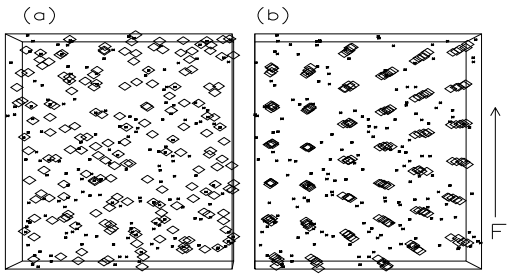}}
\caption{Pancake configuration (top view of the layers) for two typical values 
of $F$, (a) $F_c< F=0.7 <F_p$, (b) $F_p < F=2.5$. Diamonds ($\Diamond$) represent pancake 
vortices and asterisks ($*$) represent pinning centers.}
\label{fig:largenenough}
\end{figure}
\begin{figure}[htb]
\centerline{\epsfxsize=8cm \epsfbox{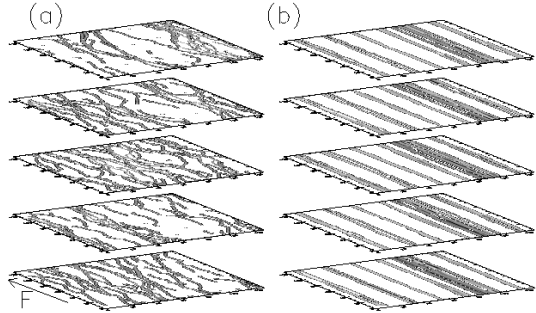}}
\caption{Pancake trajectories for two typical values of
$F$, (a) $F_c< F=0.7 < F_p$, (b) $F_p<F=2.5$, obtained plotting all the
positions of the pancakes for all time iteration steps.}
\label{fig:toosmall}
\end{figure}
\begin{figure}[htb]
\centerline{\epsfxsize=7cm \epsfbox{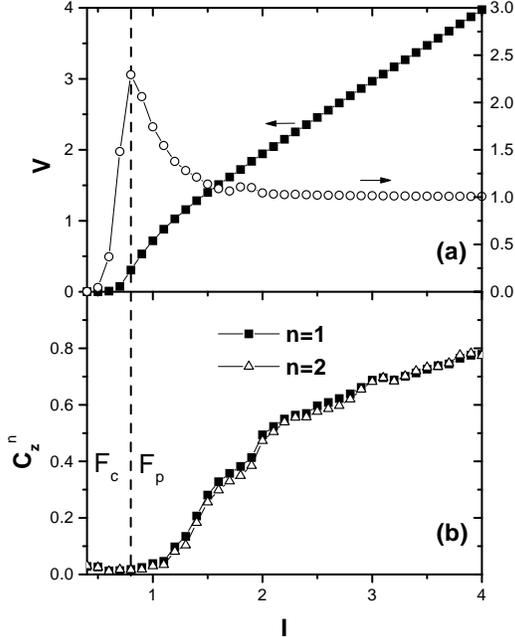}}
\caption{(a) Velocity-force curve (voltage-current characteristics), left scale,
black points, $dV/dF$ (differential-resistance), right scale, white points. (b)
Correlation parameter along c-axis $C^n_z$ for $n=1,2$.}
\label{fig:toosmall2}
\end{figure}
\section{DISCUSSION}
In a system with $N_l=5$ layers we have found that there is a clear onset of c-axis correlation 
with increasing driving force, where the pancake vortices start to align forming well defined stacks 
moving in smectic flow channels. Below this transition, in the plastic flow regime, these stacks of pancakes 
are unstable. Also, in \cite{tdgl} an enhancement of c-axis correlations with increasing drive was
observed in a bilayered system.

We have further found \cite{veremos} that the in-plane properties are in well correspondence
 with the ones obtained in 2D thin films simulations \cite{kolton}. A better understanding 
 of the effects of c-axis correlations in pancake motion on each layer can be obtained 
 by studying translational and temporal 
order in larger systems, through the analysis of the structure factor, voltage noise, and 
in-plane \cite{mingo} as well as {\it inter-plane} velocity-velocity 
correlation functions \cite{veremos}.

In conclusion we have analyzed the vortex correlation along the field
direction (c-axis) in the velocity-force characteristics at $T=0$ and found that 
above the critical current there is an onset of c-axis correlation in the 
transition between a plastic flow regime to a smectic flow regime. This transition 
coincides with the peak in the differential resistance. Experimentally, this effect 
could be studied with measurements of c-axis resistivity as a function of an applied 
current parallel to the layers \cite{lamenghi}. \newline       

We acknowledge discussions with L.N.\ Bulaevskii, P.S.\ Cornaglia, F.\ de la Cruz, Y.\ Fasano, 
M.\ Menghini and C.J.\ Olson. 
This work has been supported by a grant from ANPCYT (Argentina), 
Proy. 03-00000-01034. 
D.D.\ and A.B.K.\ acknowledge support from Fundaci\'on Antorchas
(Proy. A-13532/1-96), Conicet, CNEA and FOMEC 
(Argentina).
This work was also supported by the Director, Office of Advanced Scientific
Computing Research, Division of Mathematical, Information, and 
Computational Sciences of the U.S.\ Department of Energy under contract
number DE-AC03-76SF00098.

\end{document}